\documentclass[a4paper,fleqn]{cas-dc}

\usepackage[numbers]{natbib}

\def\tsc#1{\csdef{#1}{\textsc{\lowercase{#1}}\xspace}}
\tsc{WGM}
\tsc{QE}
\tsc{EP}
\tsc{PMS}
\tsc{BEC}
\tsc{DE}

\begin{document}
\let\WriteBookmarks\relax
\def\floatpagepagefraction{1}
\def\textpagefraction{.001}
\shorttitle{A BenchCouncil View on Benchmarking Emerging and Future Computing.}
\shortauthors{Prof. Dr. Jianfeng Zhan}

\title [mode = title]{A BenchCouncil View on Benchmarking Emerging and Future Computing. }




\author[1,3]{Jianfeng Zhan}
\cormark[1]
\fnmark[1]
\ead{zhanjianfeng@ict.ac.cn}
\ead[url]{www.benchcouncil.org/zjf.html}


\address[1]{Research Center for Advanced Computer Systems, Institute of Computing Technology, Chinese Academy of Sciences}









\begin{abstract}
The measurable properties of the artifacts or objects in the computer, management, or finance disciplines are extrinsic, not inherent --- dependent on their problem definitions and solution instantiations. Only after the instantiation can the solutions to the problem be measured. The processes of definition, instantiation, and measurement are entangled, and they have complex mutual influences. Meanwhile, the technology inertia brings instantiation bias --- trapped into a subspace or even a point at a high-dimension solution space. 
These daunting challenges, which emerging computing aggravates,  make metrology can not work for benchmark communities. It is pressing to establish independent benchmark science and engineering.

This article presents a unifying benchmark definition, a conceptual framework, and a traceable and supervised learning-based benchmarking methodology, laying the foundation for benchmark science and engineering.  I also discuss BenchCouncil's plans for emerging and future computing.  The ongoing projects include defining the challenges of intelligence, instinct, quantum computers, Metaverse, planet-scale computers,  and reformulating data centers, artificial intelligence for science, and CPU  benchmark suites. Also, BenchCouncil will collaborate with ComputerCouncil on open-source computer systems for planet-scale computing, AI for science systems, and Metaverse.  \end{abstract}

\maketitle

\section{Introduction}

\begin{figure*}
	\centering
		\includegraphics[scale=.5]{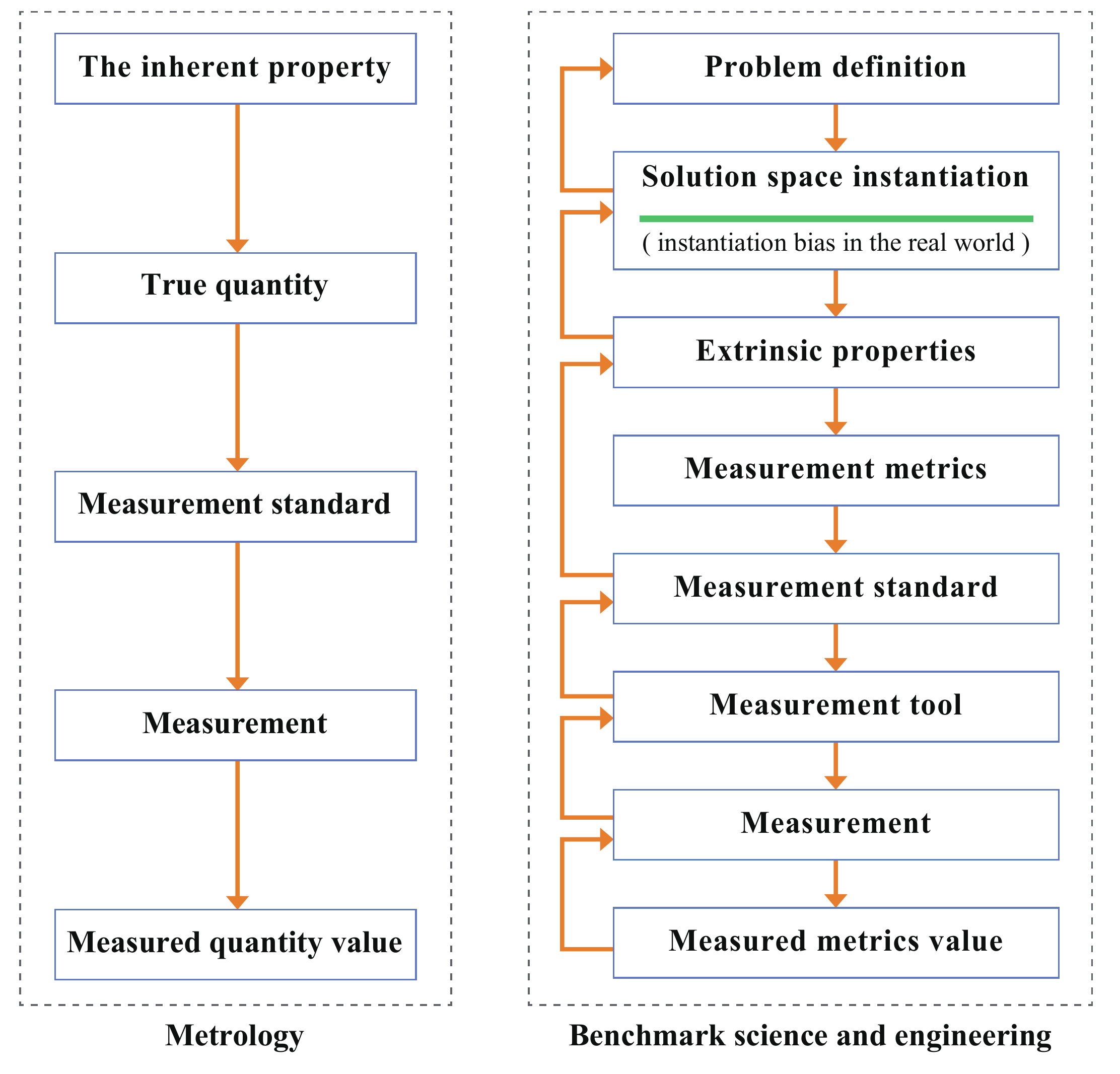}
	\caption{With respect to metrology, the benchmarking challenges --- the extrinsic properties, process entanglement, and instantiation bias --- explain why metrology can not work for the benchmark community.  First, the property of a benchmark is not inherent but dependent on its problem definition and solution instantiation. Second, only after the instantiation can the solutions to the problem be measured. The processes of definition, instantiation, and measurement are entangled, and they have complex mutual influences.  Third, instantiation introduces many biases. }
	\label{challenge_inbench}
\end{figure*}

Benchmarking is widely practiced in different disciplines without a consensus on a consistent definition. For example, in the computer science discipline, the community uses a set of workload implementations to measure CPU (processor) performances~\cite{SPECCPU2017,panda2018wait}. In machine learning, standardized data sets labeled with ground truths are used to define a data science problem~\cite{deng2009imagenet,johnson2016mimic}. In the management discipline, the industry best practices are searched and compared against different products, services, and processes~\cite{zairi1996origins, ZHAN2021100012benchmark}.  All are called benchmarks or benchmarking. In the previous work, I concluded five categories of benchmarks~\cite{ZHAN2021100012benchmark}: measurement standards, standardized data sets with defined properties, representative workloads, representative data sets, and industry best practices. 

The inconsistency or chaos results from the following fact. 
Per JCGM 200 definition, metrology is the science of measurement and its application~\cite{bipm2012international,kacker2021quantity}. Metrology measures intrinsic properties independent of an observer, like length, time, and power. There is a true quantity for each inherent property, where a probability could state the coverage interval containing the true value~\cite{bipm2012international,kacker2021quantity}. However, the measurable properties of the artifacts or objects in the computer, management, or finance disciplines are extrinsic, not inherent --- dependent on their problem definitions and solution instantiations. Unlike the processes in metrology that are linear and static, the processes of a benchmark have complex mutual influence.  Only after the instantiation can the solutions to the problem be measured.  Moreover, the definition, instantiation, and measurement processes are entangled and indivisible, which I call process entanglement. Users adhere to existing products, tools, platforms, and services, called technology inertia~\cite{zhan2022three}. The technology inertia traps the solution to a problem into a specific exploration path --- a subspace or even a point at a high-dimension solution space.  The instantiation bias impacts the measurement of the extrinsic properties. 

Our society increasingly relies upon information infrastructure with daunting complexity that dwarfs the previous computing systems, making it difficult to trace the problem definition. Instead, the biased instantiation of solutions becomes the proxy of the problems, missing the forest for the trees.   As shown in Figure~\ref{challenge_inbench}, these daunting challenges: extrinsic properties, process entanglement, and instantiation bias, result in the benchmark community's inability to reuse the metrology knowledge and the de facto isolation of benchmark communities, like computers, management, and finance, developing different methodologies, tools, and practices.   It is pressing to establish independent benchmark science and engineering.

Echoing my past call~\cite{ZHAN2021100012benchmark}, this article further builds up benchmark science and engineering. I define the benchmark from the perspective of problems and state-of-the-practice solutions. Benchmark is an explicit or implicit definition of a problem, an instantiation of a problem, an instantiation of state-of-the-practice solutions as the proxy to the problem, or a measurement standard that quantitatively measures the solution space. I propose a concise conceptual framework for the benchmark science and engineering, at the core of which is the extrinsic properties. The extrinsic property is a benchmark property that is not inherent but dependent on a problem definition and its solution instantiation. I propose a traceable
and supervised learning-based methodology to tackle the challenges of extrinsic property, process entanglement and instantiation bias. The essence of the methodology has two integrated parts: manage the traceability of the processes from the definition, instantiation, and measurement; search for the best solution through supervised learning with reference to a thoroughly-understood process from the problem definition, solution instantiation to measurement.

Also, I  discuss BenchCouncil's plan for emerging and future challenges.  The ongoing projects include defining the challenges of intelligence, instinct, quantum computers, Metaverse, planet-scale computers,  and reformulating data centers, artificial intelligence for science, and CPU  benchmark suites. Also, BenchCouncil will collaborate with ComputerCouncil~\cite{zhan2022open} on the open-source computer systems for Planet-scale computing~\cite{PSC}, AI for science~\cite{Li_SAIBench}, and Metaverse~\cite{MetaverseBench}.


The organization of this article is as follows. Section Two presents the background and challenges and explains why metrology can not work for the benchmark community. Section Three describes why emerging computing aggravates the benchmark challenges. Section Four lays the foundation for benchmark science and engineering, including the unifying definition of benchmarks, the conceptual framework, and the benchmarking methodology.  Section Five details BenchCouncil's plan. Section Six  concludes.


\section{Background and challenge: why metrology can not be directly reused for benchmark science and engineering}

\begin{figure}
	\centering
		\includegraphics[scale=.5]{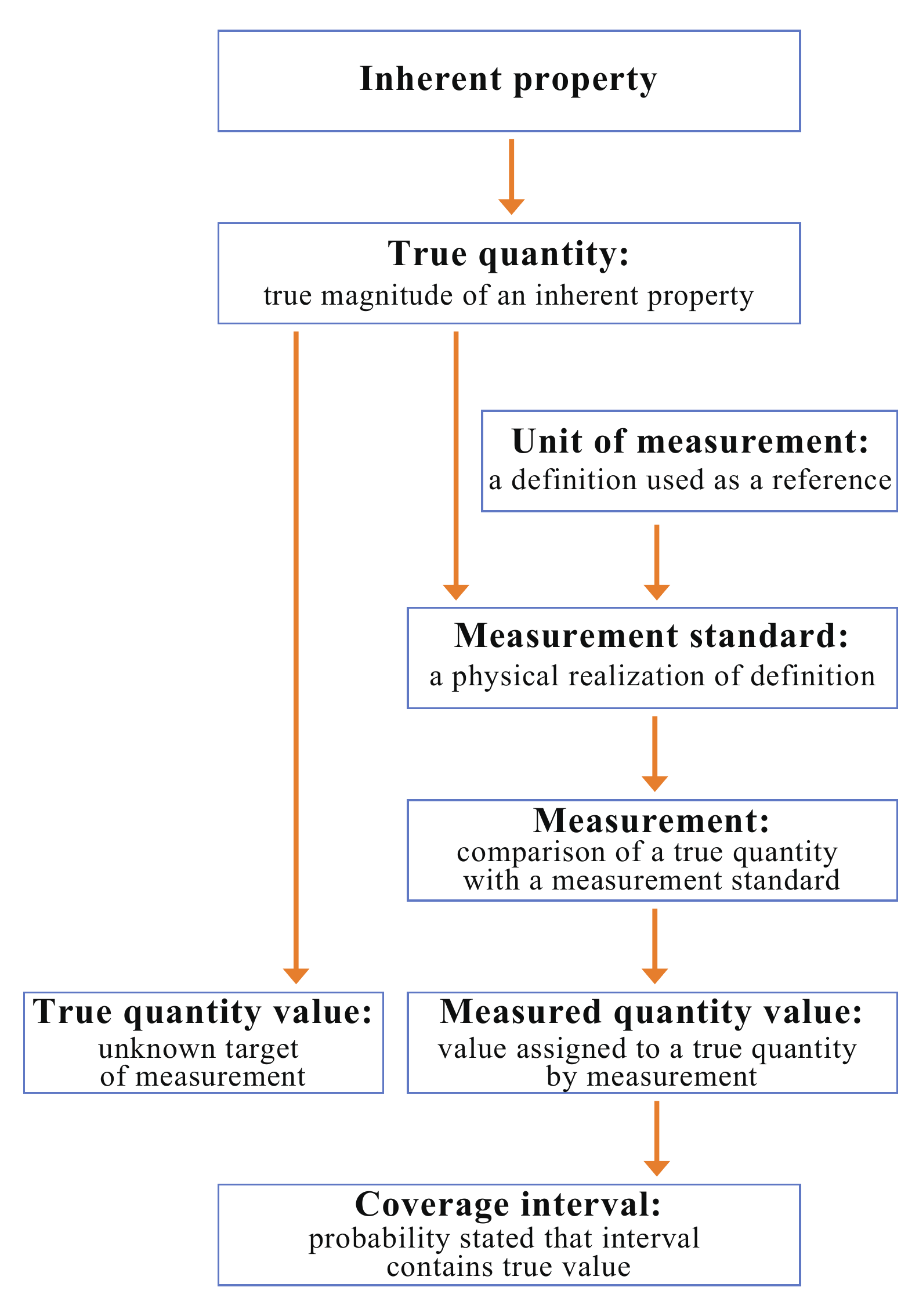}
	\caption{A simple but systematic metrology conceptual framework is used to clarify why metrology can not be directly reused for benchmark science and engineering. Some concepts are defined by myself, while the other concepts are reused or modified from ~\cite{bipm2012international, kacker2021quantity}. Only necessary metrology concepts are reserved to keep concise.  }
	\label{metro-fram}
\end{figure}

In this section, I first introduce the metrology concepts as background and then present the benchmarking challenges and why metrology can not work in the benchmark community.


\subsection {Background: metrology concepts}~\label{metrology_concept}

As shown in Figure~\ref{metro-fram}, I provide a simple but systematic metrology concept framework to clarify why metrology can not be directly reused for establishing benchmark science and engineering. I present and modify most of those concepts from~\cite{bipm2012international, kacker2021quantity}. But I define some concepts to emphasize why metrology can not work for the benchmark community, e.g., inherent properties. To keep concise, I only stay with necessary metrology concepts. 

The inherent property is a property of a phenomenon, body, or substance that is independent of an observer, e.g., length and energy~\cite{bipm2012international}. The inherent property can have various magnitudes. True quantity is the magnitude of an inherent property of an individual phenomenon, body, or substance that is independent of an observer, e.g., the radius of a given circle, the kinetic energy of an identified particle in a given system~\cite{kacker2021quantity,bipm2012international}. 
Unit of measurement~\cite{kacker2021quantity} is a definition and its physical realization, used as a reference to assign a value to a true quantity. Measurement standard~\cite{kacker2021quantity} is a physical realization of a unit of measurement, with a stated quantity value and associated measurement uncertainty. 

Measurement~\cite{kacker2021quantity} is a process of comparing a 
true quantity with a measurement standard to assign the true quantity one or more quantity values that are traceable to a unit of measurement. A quantity value obtained by the measurement is referred to as a measured quantity value~\cite{bipm2012international}.  True quantity value~\cite{bipm2012international} is a quantity value consistent with the definition of a quantity, which is an unknown measurement target~\cite{bipm2012international}. A coverage probability~\cite{bipm2012international} is a probability that a specified coverage interval contains the true quantity value.


\subsection{The benchmarking challenges: extrinsic properties, process entanglement, and instantiation bias} \label{challenge_extrinsic_instantiation}


In the previous work~\cite{ZHAN2021100012benchmark}, I have noticed the differences in properties of the artifacts or objects in the computer, management, or finance disciplines from those classical ones,  like length, time, and power. The properties of the artifacts or objects in the computer, management, or finance disciplines are extrinsic, dependent on their problem definitions and solution instantiations. Instead, the classical properties like time and length are inherent, independent of the observers. 
From a concept perspective, it is easy to say there are three essential processes: definition, instantiation, and measurement. However, a problem definition is abstract; only after the instantiation can the solutions to the problem be measured. Moreover, the definition, instantiation, and measurement processes are entangled and indivisible, which I call process entanglement.  Only by fully understanding the side effect of the extrinsic properties and process entanglement can we avoid many traps. I elaborate on this viewpoint from different perspectives. Before proposing the conceptual framework for benchmark science and engineering (I defer it to Section~\ref{conceptual_framework}), I stay with the metrology concepts in Section~\ref{metrology_concept} to depict the challenges.

A subtle change in the definitions of a problem may lead to wildly varied solutions and significantly different quantity value. I take the classical matrix multiplication problem~\cite{williams2012multiplying, blalock2021multiplying} as an example. Blalock et al.~\cite{blalock2021multiplying} reformulate the classical matrix multiplication problem as follows. The following reformulation is cited from~\cite{blalock2021multiplying}. A and B are two matrices. A is $R^{NxD}$ and B is $R^{DxM}$, $N>>D>=M$. Given a computation time budget $\tau$, the task constructs three functions g(·), h(·), and f(·), along with constants $\alpha$
and $\beta$, such that
\begin{equation}
    \label{formula-1}
    ||\alpha f(g(A),h(B))+ \beta - AB||_{F}< \epsilon(\tau)||AB||_{F}
\end{equation}
for as small an error $\epsilon(\tau)$ possible.  For this reformulated problem, they introduce a learning-based algorithm  that greatly outperforms existing methods~\cite{blalock2021multiplying}. This is a typical example of a subtle change in the definitions of a problem leading to wildly varied solutions and significantly different measured quantity values.

Furthermore, the solutions instantiations  also interplay with each other and finally impact measured quantity values. The obvious example is deep learning. Algorithms and neural network architectures play a significant role. The algorithm and hardware implementation, like different precision, e.g., single-precision, double-precision or mixed precision, impact the learning dynamics. 
Even for the same system with different scales, the interactions among system size and minibatch size significantly impact the  measured quantity values like time-to-quality --- the training time to achieve the state-of-the-quality --- and FLOPS (the computation overhead)~\cite{jiang2020hpc, goyal2017accurate,you2018imagenet, mattson2020mlperf,tang2021aibench}. 

The processes in metrology are linear and static. However, as shown in Figure~\ref{challenge_inbench}, the processes of definition, instantiation, and measurement of a benchmark are entangled, having complex mutual influences.  The subtle difference in a problem definition will lead to a wildly varied solution, and its instantiation finally significantly impacts the measured quantity value.  The solution instantiation provides the basis for measurement tools, as the latter often use state-of-the-practice instantiation, which affects the measured quantity value. The measured quantity values provide hints on searching for the best instantiation in the solution space.

Moreover, the instantiation introduces many biases, which I call instantiation bias.  For example, in the computer system and architecture disciplines,   
Wang et al.~\cite{wang2021wpc,wang2022wpc} found that merely conducting ISA-independent (ISA is short for instruction set architecture), microarchitecture-dependent or microarchitecture-independent workload characterization (a form of measurement) will lead to misleading or erroneous conclusions. These significant differences in measured quantity values resulted from the solution instantiations at different levels themselves. Before performing ISA-independent or microarchitecture-dependent/independent workload characterization, the necessary step is instantiating a computer workload benchmark on a specific microarchitecture, a particular instruction set architecture, or an intermediate representation (very close to the source code), respectively. The community opts for the widely used ISA, IR (intermediate representation) for instantiation. 

Matsuoka et al. also found the implementation of biases and complexity traps~\cite{matsuoka2022preparing} in the instantiation process: on the one hand, any implementation of a computer workload benchmark entails multiple implicit biases towards algorithms, programming languages, data layouts, and parallelization approaches; on the other hand,  the benchmarks, abstracted from large or legacy scientific codes and tuned for previous computer architectures, trap the co-design participants into considering only similar architectures. 

Other observations are from the data sets, which many communities like machine learning use to explicitly or implicitly define a problem. It is prohibitively costly to build a representative and fidelity data set that can capture real-world characteristics. Hence, in reality, the goal often degrades to a workable data set. For example, for ImageNet~\cite{deng2009imagenet}, it is easy to collect familiar animal and plant pictures, while the rare ones are difficult to obtain. Considering the data set is the cornerstone of many challenges like auto-driving and automatic medical diagnosis, this far-fetched methodology has many hidden flaws and risks.

\section{Emerging computing aggravates the challenges\protect\footnote{This section is written based on an unpublished technique report~\cite{zhan2019benchcouncil}, of which I am the lead author.}}

Modern society is digitized, increasingly relying upon information infrastructure. The information infrastructure consists of massive Internet of Things (IoT), edge devices, data centers, and high-performance computers. Those systems collaborate to handle big data, train AI models and provide Internet services augmented by AI inference for huge end-users with guaranteed quality of services. From a benchmarking perspective, emerging computing like Big Data, AI, and Internet Services are significantly different from the traditional workloads characterized by SPECCPU (desktop workloads)~\cite{SPECCPU2017}, TPC-C~\cite{tpcc}, TPC-Web (Traditional web services)~\cite{tpc-w}, and HPL (high-performance computing)~\cite{petitet2004hpl} benchmarks, raising serious challenges. 

The first challenge is fragmentation. There are substantial fragmented application scenarios, a marked departure from the past~\cite{zhan2019benchcouncil}. For example,  hundreds or even thousands of ad-hoc big data solutions, termed NoSQL or NewSQL, are proposed to handle different application scenarios. For AI, the same observation holds. There are tens or even hundreds of organizations that are developing AI training or inference chips to tackle their challenges in different application scenarios, respectively ~\cite{mattson2020mlperf,reddi2020mlperf}. 

The second challenge is de facto isolation. Internet service provider giants own and treat real-world data sets and workloads or even AI models as first-class confidential issues. The treasure is hidden in data centers and isolated between academia and industry, or even among different providers~\cite{gao2019aibench}. The dire situation poses a huge obstacle for our communities towards developing an open and mature research field~\cite{gao2019aibench}.

The third challenge is the complexity of collaboration: HPC systems, data centers, edge, and IoT devices collaboratively handle the challenges; In the collaborations, different distributions of data sets, workloads, machine learning, or AI models may substantially affect the system's behaviors; the interaction patterns among IoT, edge, and data centers changes fast, embodying different architecture. 

The fourth challenge originates from service-based architecture. On the one hand, the software-as-a-service (SaaS) development and deploy model makes the workloads change very fast (so-called workload churn) ~\cite{barroso2009datacenter}, and it is not scalable or even impossible to create a new benchmark or proxy for every possible workload ~\cite{gao2018motif}. On the other hand, modern Internet services adopt a microservice-based architecture, often consisting of various modules with long and complex execution paths across different data centers. As the worst-case performance (tail latency) ~\cite{dean2013tail} does matter, the micro-service-based architecture also poses a severe challenge to benchmarking~\cite{gao2019aibench,gao2021aibench}.

The final but not least challenge is the stochastic nature of AI. AI techniques are widely used to augment modern products or Internet services. The nature of AI is stochastic, allowing multiple different but equally valid solutions ~\cite{mattson2020mlperf}. Many factors manifest the uncertainties of AI, e.g., the adverse effect of lower-precision optimization on the quality of the final model, the impact of scaling training on time-to-quality, and run-to-run variation in terms of epochs-to-quality~\cite{mattson2020mlperf}. However, the benchmarks mandate being repeatable (the same team) and reproducible (different teams). This conflict raises serious challenges.

Emerging computing aggravates the benchmarking challenges discussed in Section~\ref{challenge_extrinsic_instantiation}. First, it is difficult to trace the original problem definition, that is, the target to be achieved. Second, taking the instantiation of solutions as the proxy for the problem aggravates the instantiation bias and makes the community further trapped in the specific solutions.

\section{Building up benchmark science and engineering}

This section proposes the unifying definitions of benchmarks, the conceptual framework, and the benchmarking methodology, which lays the foundation for benchmark science and engineering.

\subsection{The unifying definition of benchmarks}~\label{unifying_definition} 

Previously, I concluded five categories of benchmarks~\cite{ZHAN2021100012benchmark}: measurement standards, standardized data sets with defined properties, representative workloads, representative data sets, and industry best practices. 
In this section, I give a simple and unifying definition to cover five categories of benchmarks and reveal their essence. 
A benchmark is an explicit or implicit definition of a problem, an instantiation of a problem, an instantiation of state-of-the-practice solutions as the proxy to the problem, or a measurement standard that quantitatively measures the solution space. 


A benchmark has three essential processes, some of which often be omitted or implicitly stated in practice: definition, instantiation, and measurement. I explain the process of definition and instantiation from various perspectives in the rest paragraphs of Section~\ref{unifying_definition}.   I leave the discussion of the process of measurement in Sections ~\ref{conceptual_framework}, ~\ref{benchmark_methodology}.


\subsubsection{Definition}

The first process is the definition. Defining a problem explicitly or implicitly is the fundamental role that a benchmark could play in almost all disciplines. Only after clearly defining a problem can we figure out the solutions and compare them against the others. For example, Alan Turing, in 1950~\cite{turing_test} formulated the problem of what intelligence is as an imitation game: the game tests whether an interrogator can distinguish a machine's ability from a human. Turing's problem definition inspires several generations to explore the solutions to achieve intelligence.

There are many ways to define a problem, e.g., using a natural language or mathematics.  From an accuracy perspective, mathematically defining the problem is a better choice. Unfortunately, many problems can not be accurately depicted in this way.  

The NAS parallel benchmarks~\cite{bailey2011parallel} claimed that the common requirements should be specified in a paper-and-pencil approach~\cite{zhan2019benchcouncil}.
A paper-and-pencil approach is a vague description -- It can be mathematical, textual, or even visually.  In the computer science discipline, this approach is well-practiced in the database community but not adopted in the computer architecture community. 

Shun et al.~\cite{shun2012brief} advocated a methodology to build benchmarks using problem definitions, and they created the problem-based benchmark suite (PBBS). PBBS is a set of benchmarks for comparing parallel algorithmic approaches, parallel programming language styles, and machine architectures across a broad set of problems. Specifically, a problem-based benchmark mandates a problem specification and a set of input distributions while not detailing the requirements in terms of algorithmic approach, programming language, or machine architecture ~\cite{shun2012brief}.


\subsubsection{Instantiation}

The second process is instantiating a problem or instantiating state-of-the-practice solutions as the proxy to the problem or challenge. As a replacement or complement, these are two different ways.  First, an instantiation of the problem is used. For example, a data set is often used to instantiate a problem in the machine learning community.  Second, an instantiation of state-of-the-practice solutions is used as the proxy for the problem. For example, the computer architecture community provides state-of-the-practice implementations of a group of computer workloads like SPECCPU~\cite{henning2000spec,panda2018wait}. SPECCPU is a proxy to the problems. 

There are two reasons for this replacement or complement. First, as a replacement, it serves as the proxy for the problem that is too difficult to define. Second, as a complement, the instantiation brings enriched and necessary details that set more specific problem settings. Each instantiation is a subspace or point --which is often state-of-the-practice-- in the solution space to the problem, e.g., using source code or binary code, which brings instantiation bias.

\subsection{The conceptual framework of benchmark science and engineering}\label{conceptual_framework}

 \begin{figure}
	\centering
		\includegraphics[scale=.5]{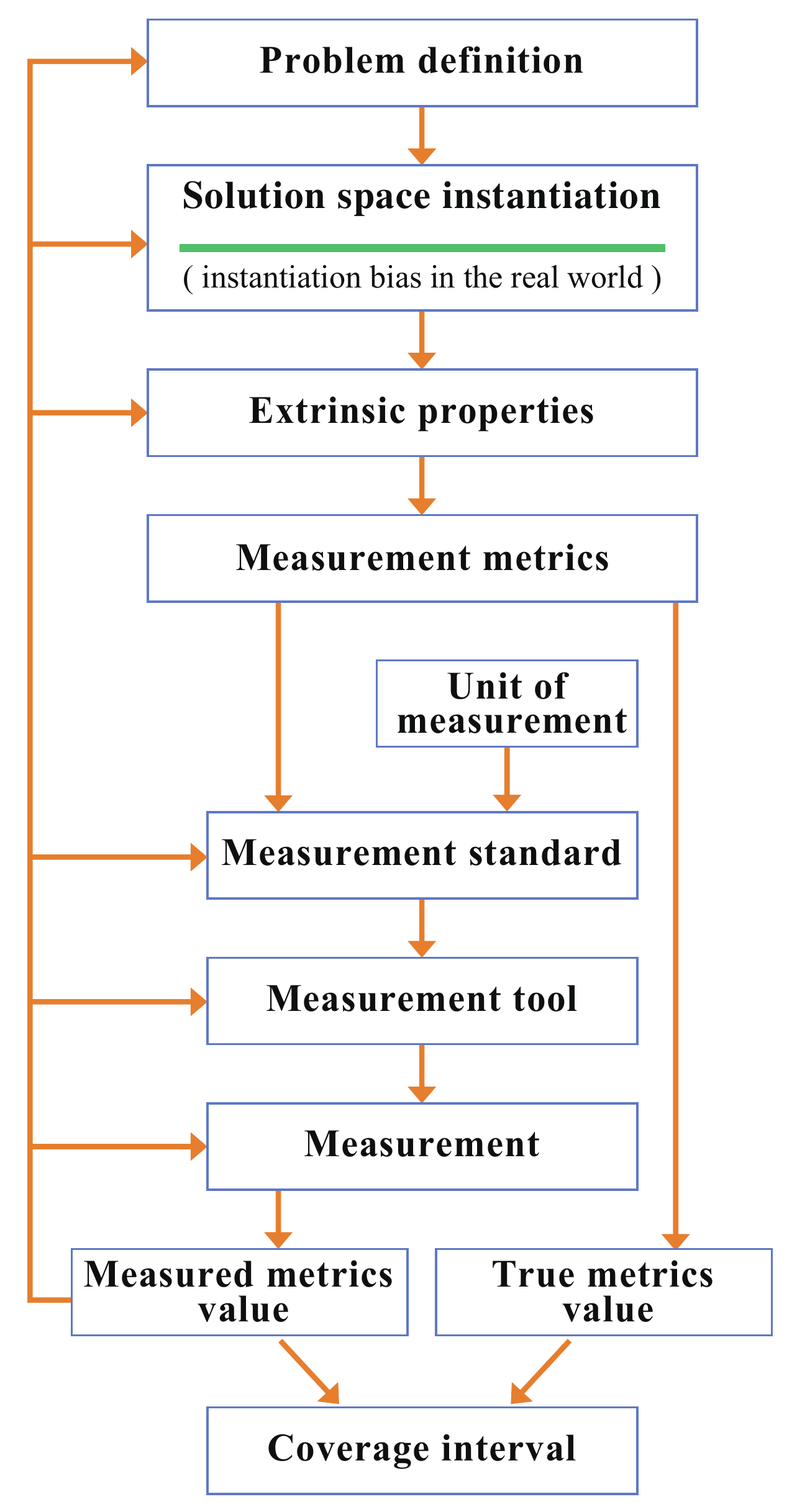}
	\caption{The conceptual framework of benchmark science and engineering.}
	\label{benchmark_conceptual_framework}
\end{figure}
As shown in Figure~\ref{benchmark_conceptual_framework}, I propose  the conceptual framework of benchmark science and engineering.
The extrinsic property is a benchmark property that depends on a problem definition and its solution instantiation. The extrinsic property can have various magnitudes. Measurement metrics are the magnitude of a benchmark's extrinsic property, which depends on a problem definition and solution instantiations. Unit of measurement~\cite{kacker2021quantity} is a definition and its realization, used as a reference to assign a value to a measurement metric. Measurement standard~\cite{kacker2021quantity} is a realization of a unit of measurement, with a stated metric value, associated measurement uncertainty, and a repeatable (the same team) and reproducible (different teams) measurement methodology. The measurement tool implements a measurement standard that can be calibrated and traceable. Traceability~\cite{bipm2012international} is a property of a measurement result whereby the result can be related to a reference through a documented unbroken chain of calibrations, each contributing to the measurement uncertainty. The measurement tools should be open-sourced and can be replicated by different teams. 

Measurement~\cite{kacker2021quantity} is a process of comparing a 
measurement metric with a measurement standard to assign one or more measured values to a measurement metric that are traceable to a unit of measurement. A value obtained by the measurement is referred to as a measured metric value~\cite{bipm2012international}.  True metric value~\cite{bipm2012international} is a value consistent with the definition of a measurement metric, which is an unknown measurement target~\cite{bipm2012international}. A coverage probability~\cite{bipm2012international} is a probability that the true metric value is contained within a specified coverage interval.

\subsection{The traceable and supervised-learning based benchmarking methodology}~\label{benchmark_methodology}



\begin{figure*}
	\centering
		\includegraphics[scale=.45]{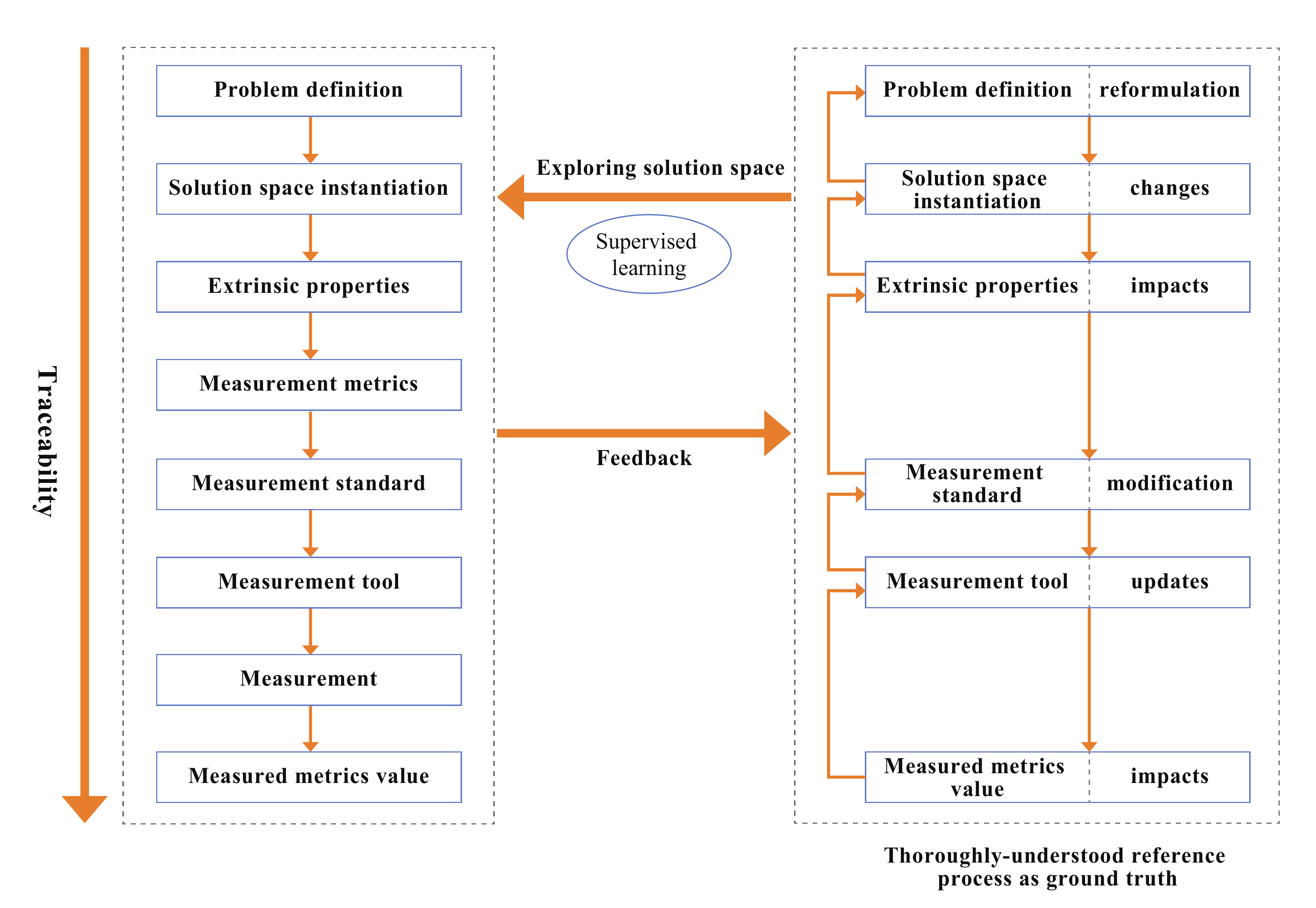}
	\caption{A traceable and supervised learning-based benchmarking methodology to tackle the challenges of extrinsic property, process entanglement, and instantiation bias.}
	\label{method}
\end{figure*}

A benchmark has no inherent properties, and its extrinsic property is dependent on its problem definition and solution instantiation. Meanwhile, the processes of definition, instantiation, and measurement are entangled, and they have complex mutual influences. 

I propose a traceable methodology to tackle the above challenge, at the core of which is to manage the traceability of the processes from the definition, instantiation, and measurement.  
Figure~\ref{method} shows that problem definition, solution instantiation, extrinsic properties, measurement standard, measurement tool,  and measured metrics value have complex mutual influence.  The problem definition is the origin of this relationship. No other below entities, like solution instantiation, can impact the problem definition directly. Still, solution instantiations may provide clues for the subtle change of the problem definition, affecting the other entities significantly. At the top level, I suggest a formal definition of problems and tracing the relationship among the different subtle definitions of problems.  For many state-of-the-practice benchmarks, the definition process is omitted. It should regularly keep an eye on revisiting the process, or else the outdated instantiation will be a trap.  The solution instantiation provides the basis for measurement tools. It is mandatory to search for state-of-the-art or state-of-the-practice solutions and implement them in the measurement tool. 





There is an explosion from the problem definition to solution instantiations. Put in other words, the lower level has more state space~\cite{ZHAN2021100012benchmark}. For example, there is increasing state space in a computer workload benchmark, from a mathematical problem definition, an algorithm, an intermediate representation, an ISA-specific representation, to a micro-architectural representation. 
The technology inertia traps the solution into a specific exploration path --- a subspace or even a point at a high-dimension solution space, called instantiation bias.  The instantiation bias impacts the measurement of the extrinsic properties.  Also, an unguided exploration may drift away from optimized solutions. 

\begin{figure*}
	\centering
		\includegraphics[scale=0.38]{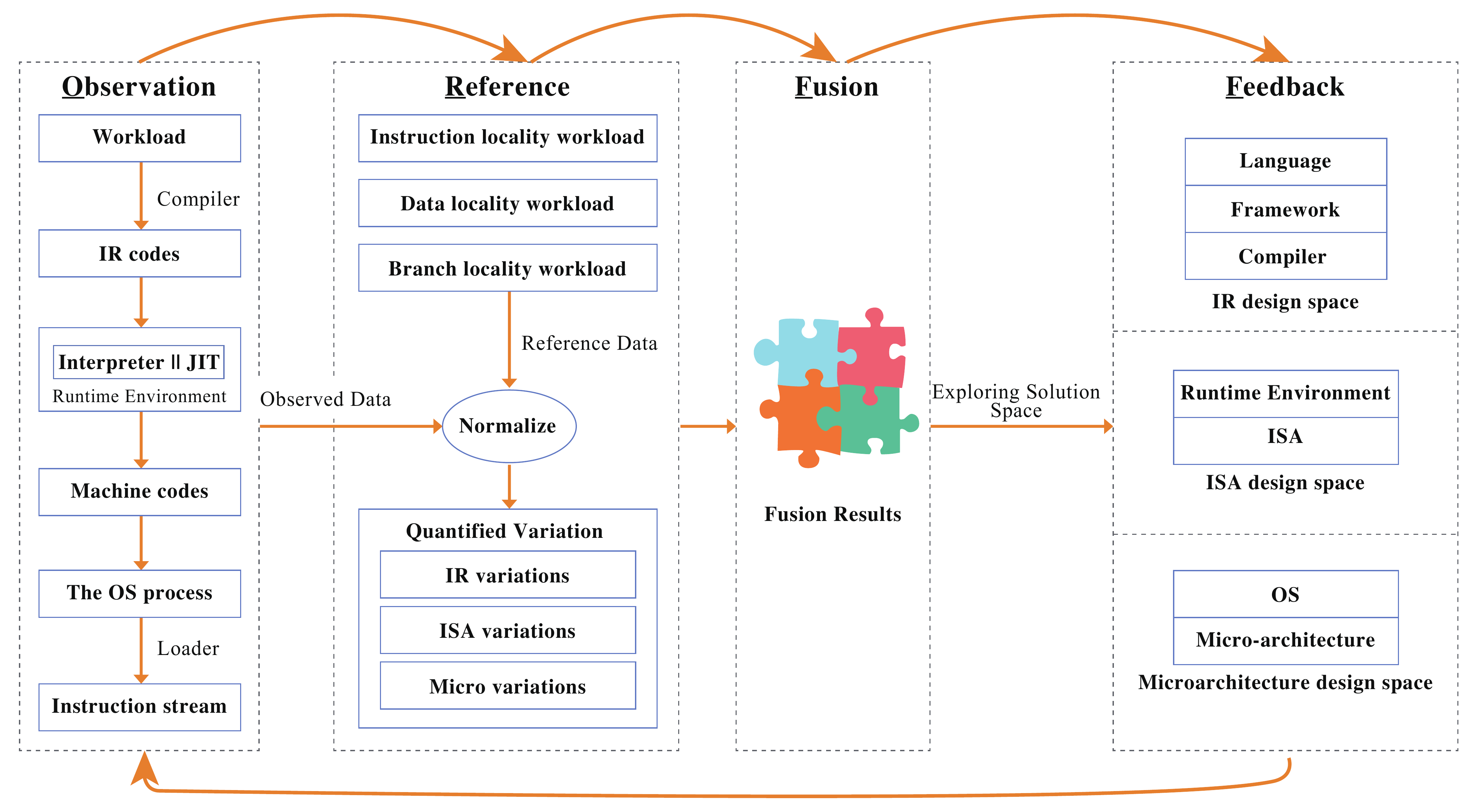}
	\caption{The application of the traceable and supervision learning-based benchmarking methodology in computer architecture. This figure is cited from~\cite{wang2022wpc} with the permission of the authors.}
	\label{application_benchmark_conceptual_framework}
\end{figure*}

I propose the supervised learning-based methodology to tackle the challenge of instantiation bias. Supervised learning is a branch of machine learning, referring to a class of algorithms that train a predictive model using labeled data with known outcomes. Figure~\ref{method} shows the thoroughly-understood process from the problem definition, solution space instantiation, extrinsic properties, measurement standard, measurement tool,  to measured metrics value, standing as a ground truth. From this ground truth, it is easy to learn how the change of the top entities impacts the below. For example, if the problem is reformulated, the solution instantiation changes accordingly. Finally, the measured metrics values are significantly affected. The benchmark plays the role of connecting the problem with its solution space. By exploring the solution space and observing the effect of its change on the measured metrics values, it is possible to search for the best solution. This search process could leverage state-of-the-art deep learning techniques. Of course, this learning dynamic will be very complex. 
Figure~\ref{application_benchmark_conceptual_framework} shows an example on how to use this methodology in computer architecture.



\subsection{Re-interpret five categories of benchmarks}

I use the benchmark definition proposed in~\ref{unifying_definition} to re-interpret five categories of benchmarks  defined in~\cite{ZHAN2021100012benchmark}. 

The first category of the benchmark is a measurement standard used to measure the solution space to the problem. I use the Linpack benchmark --- an example from high-performance computing --- to explain this category of benchmarks. The Linpack benchmark~\cite{dongarra2003linpack} is widely used to report the performance of a high-performance computer.  The problem definition of Linpack is a linear system of equations of an order n: $Ax=b$. The solution uses the LU factorization with partial pivoting. The measurement metrics are the floating-point operations count of the solving algorithm, which is ($2*n^{3}/3 + 2*n^{2}$) operations, and the execution time of running the benchmark.  HPL is one of the reference implementations of the measurement tool used to evaluate against different high-performance computer implementations (solutions). The measurement standard also details the reproducible and repeatable measurement methodology to compare against other solutions: users must report a residual for the accuracy of the solution with  $||Ax-b||/(||A||||x||)$.  The TOP500 list reports the measured metrics values. The measurement metrics are higly depends on its problem definition and solution instantiations.

The second category of benchmarks is the representative workloads that run on the systems under measurement~\cite{ZHAN2021100012benchmark}. The representative workloads are the definition of the problem or an instantiation of state-of-the-practice solutions as the proxy to the problem.  The problem-based benchmark suite (PBBS)~\cite{shun2012brief}, TPC-C~\cite{tpcc}, TPC-Web (Traditional web services)~\cite{tpc-w} are typical problem definition examples. They also provide the instantiation of state-of-the-practice solutions as the measurement tool.  Without explicit problem definition, SPECCPU (desktop workloads)~\cite{SPECCPU2017}, BigDataBench~\cite{wang2014bigdatabench}, BigBench~\cite{ghazal2013bigbench}, AIBench~\cite{tang2021aibench,gao2021aibench} and MLPerf~\cite{mattson2020mlperf} are the instantiations of  state-of-the-practice solutions and they serves as the proxy to the problem.  

The third category of the benchmarks is the implicit definition of the problem using a standardized data set. The standardized data set represents a real-world data science problem with defined properties, some of which have ground truth~\cite{ZHAN2021100012benchmark,automl2021git}. ImageNet~\cite{deng2009imagenet} (deep learning benchmark) and MIMIC-III~\cite{johnson2016mimic} (critical care benchmark) are typical examples. 

The fourth category of benchmarks is a representative data set, used as a reference~\cite{ZHAN2021100012benchmark}. This category of benchmarks is an instantiation of a problem. For example, an index (statistical measure) calculated from a representative set of underlying data and used as a reference for financial instruments or contracts~\cite{iosco2013finben} is a benchmark in finance. The London Interbank Offered Rate (Libor) and the Euro Interbank Offered Rate are well-known financial benchmarks ~\cite{iosco2013finben,ZHAN2021100012benchmark}.

The fifth category of benchmarks is the industry best practices in diverse domains~\cite{ZHAN2021100012benchmark}.  Benchmarking is continuously searching the industry best practices with superior performance and measuring products, services, and processes against them~\cite{zairi1996origins, ZHAN2021100012benchmark}. The industry best practices are instantiations of the state-of-the-practice solutions to the problem or grand challenge.

\begin{figure*}
	\centering
		\includegraphics[scale=.55]{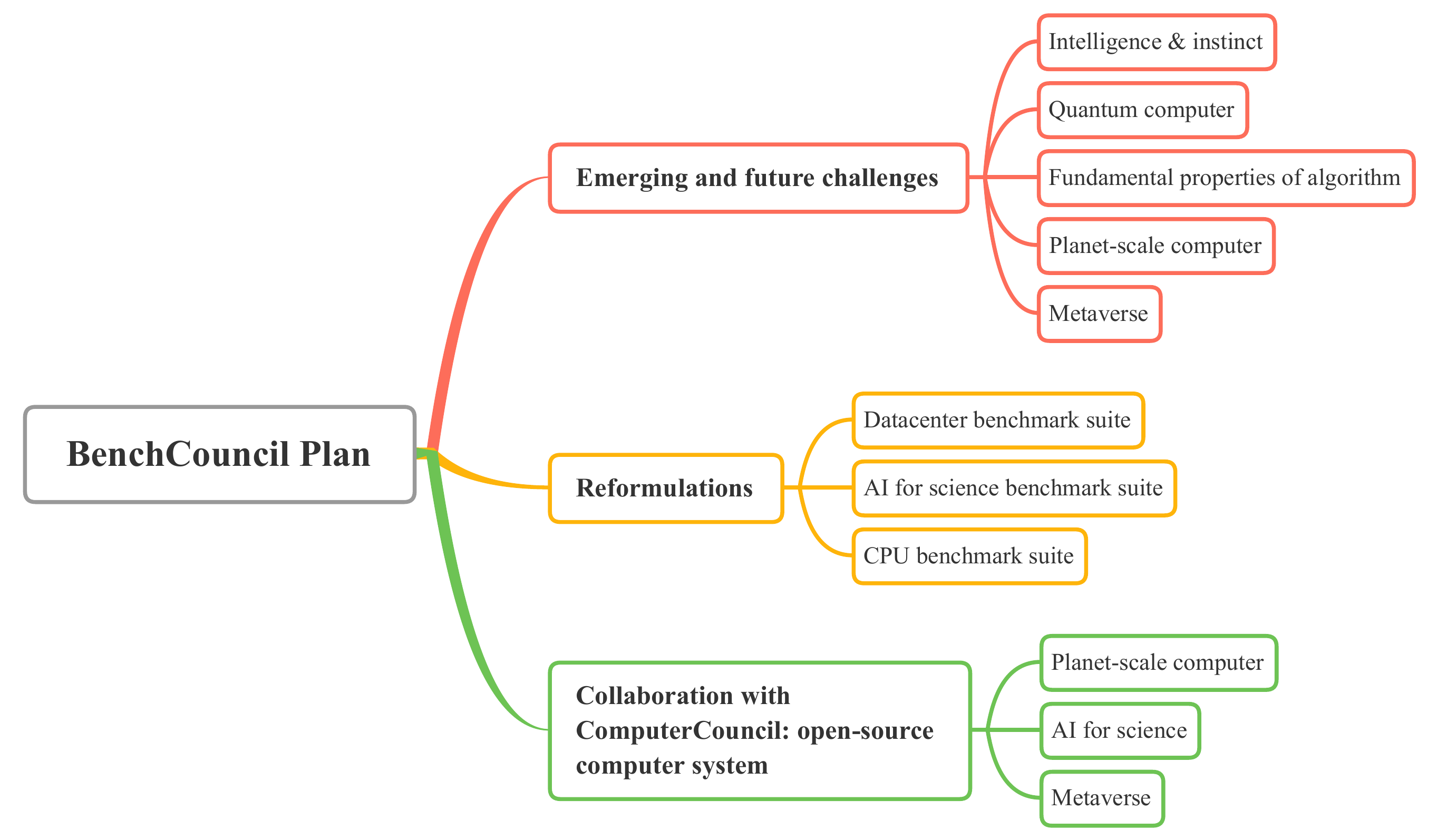}
	\caption{BenchCouncil's plan on defining the challenges of emerging and future computing and the collaboration with ComputerCouncil on the open-source computer systems. }
	\label{bc_plan}
\end{figure*}

\section{BenchCouncil's plan on emerging and future computing}~\label{benchcouncil_plan}

Figure~\ref{bc_plan} presents BenchCouncil's plan for defining the challenges of emerging and future computing and collaborating with ComputerCouncil on the open-source computer systems. First, I introduce BenchCouncil's plan for emerging and future challenges.

First, what is intelligence? What is instinct? What is the distinction between intelligence and instinct? The pre-trained language models, such as BERT and GPT-3, seem to outperform the capability beyond the Turing test~\cite{manning2022human}. Many previous works have reformulated the problem of what intelligence is~\cite{french2000turing, hoffmann2022ai}. It is necessary to revisit the processes from the intelligence problem definition, solution instantiations to measurement. For example, there are many ways to solve these challenges to somewhat extent, including traditional machine learning, deep learning, and brain-inspired computing. Letting them compete in the same arena is essential. According to~\cite{instinct}, instinct is an inborn impulse or motivation to action typically performed in response to specific external stimuli. But how do we distinguish intelligence from instinct? What are the differences between the octopus, birds, apes, and ants' behaviors? Are they intelligence or instinct?
 
 Second, quantum computers emerge as a new computational paradigm with unprecedented capability~\cite{tomesh2022supermarq}; what are the problems or grand challenges which the quantum computers do the best? How do state-of-the-practice computers compete against quantum computers in handling different or overlapping domains of problems or grand challenges? 
 The community must ponder this fundamental issue before delving into different levels of instantiations of solutions. 
 
 Third, the computer algorithms almost govern the running of our society. It is pressing to think, specify and test what fundamental properties an algorithm must have before being embedded in our society. Think about Twitter and Facebook's impact on the election in many vote-based democratic societies. It is vital to propose benchmarks against those algorithms before putting them into practice. 
 
 Fourth, information infrastructure becomes the cornerstone of our society~\cite{zhan2022open}, and many fundamental applications like medical emergency management and smart cities applications rely upon planet-scale distributed systems consisting of massive Internet of Things (IoT) devices, edges and data centers, which I call planet-scale computers~\cite{PSC}. Different distributions of data sets, workloads, machine learning, or AI models may substantially affect the system's behaviors, and the system architectures are undergoing fast evolution regarding the interactions among IoT, edge, and data centers~\cite{zhan2019benchcouncil}. How can the community propose benchmarks for those ultra-scale emerging and future applications~\cite{gao2021aibench}? 
 
Metaverse is an umbrella term. It is predicted to be a fresh way for people to immersively access the Internet, interact with each other or digital avatars in the cyberworld, and manage digital assets.  Though many industry giants are pushing toward those goals, the process itself is in a Cambrian explosion of different things in the forms of concepts, prototypes, products, or services. It is pressing to propose a benchmark suite to define the Metaverse problem or challenge, explore and evaluate state-of-the-art and state-of-the-practice solutions~\cite{MetaverseBench}.

 Many old problems need reformulation. For example, the Berkely multidisciplinary groups proposed to use thirteen “Dwarfs”~\cite{asanovic2006landscape} --- A dwarf is an algorithmic method that captures a pattern of computation and communication --- to design and evaluate parallel programming models and architectures.  When AI has seen as a new dawn in the traditional and emerging scientific area, how to reformulate those problems~\cite{Li_SAIBench}? 
 
Datacenters have become the fundamental infrastructure of modern society. There are substantial fragmented application scenarios in big data, AI, and Internet service areas, a marked departure from the past~\cite{zhan2019benchcouncil}. Virtual technologies like containers are widely used as resource management and performance isolation facilities.  However, the current BenchCouncil benchmark suites like BigDataBench~\cite{wang2014bigdatabench} and AIBench~\cite{gao2021aibench,tang2021aibench} are fragmented without providing a full-picture definition of the problems or challenges in data centers.  Moreover, a lack of simple but elegant abstractions prevents achieving both efficiency and general purpose~\cite{zhan2019benchcouncil}.  For example,  hundreds or even thousands of ad-hoc NoSQL or NewSQL solutions are proposed to handle different big data application scenarios~\cite{zhan2019benchcouncil}. Contrasted, the relation algebra is generalized for the database theory and practice, and any complex query can be written using five primitives like select, project, product, union, and difference~\cite{codd2002relational}.  Though domain-specific software and hardware co-design is promising ~\cite{hennessy2018new}, the lack of simple but unified abstractions has two side effects~\cite{zhan2019benchcouncil}: it is prohibitively costly to build an ad-hoc solution;  single-purpose systems and architectures are structural obstacles to resource sharing.  Proposing simple but elegant abstractions is an integrated part of managing the traceability of the process from the problem definition to solution instantiation.

The CPU benchmark suite like SPECCPU~\cite{henning2000spec,panda2018wait} advanced the evolution of different processor architectures. However, the SPEC CPU is an instantiation of state-of-the-practice solutions as the proxy to the problem, severely biased towards market-dominant CPU architecture, high-performance languages like C, and high-performance computing and desktop workloads. The BENCHCPU project~\cite{benchcpu} will propose a new CPU benchmark suite. 

\subsection{The collaboration with ComputerCouncil}~\label{computercouncil_collaboration}


As a non-profit international organization, the International Opensource Computer Council (ComputerCouncil)  mission is to unite the science and technology community to tackle the challenges of information technology decoupling~\cite{comcouncil}.  ComputerCouncil initiates the open-source computer system (OSCS) initiative to tackle the challenges of IT decoupling.

ComputerCouncil will choose three emerging areas: planet-scale computers --- planet-scale distributed systems and applications built on IoTs, edges, and datacenters~\cite{PSC}, AI for science~\cite{Li_SAIBench}, and Metaverse~\cite{MetaverseBench} as the initial targets of the OSCS initiative. BenchCouncil will cooperate with ComputerCouncil: the former focuses on the benchmarks, while the latter concentrates on the open-source computer systems for the three emerging areas.

\section{Conclusion}

This article concluded benchmarking challenges as extrinsic properties, process entanglement, and instantiation bias. The measurable properties of a benchmark are not inherent but dependent on their problem definitions and solution instantiations. The processes of definition, instantiation, and measurement are entangled and have complex mutual influences. The technology inertia leads to a specific exploration path --a subspace or even a point at a high-dimension design space.  Those challenges make metrology can not work for benchmark communities and call for independent benchmark science and engineering.   

I  proposed a unified benchmark definition, a conceptual framework, and a traceable and supervised learning-based benchmarking methodology, laying the foundation for benchmark science and engineering. 
Benchmark is an explicit or implicit definition of a problem, an instantiation of a problem, an instantiation of state-of-the-practice solutions as the proxy to the problem, or a measurement standard that quantitatively measures the solution space. At the core of the conceptual framework, the extrinsic property is a benchmark property dependent on a problem definition and its solution instantiation. The essence of the proposed benchmarking methodology has two integrated parts: manage the traceability of the processes from the definition, instantiation, and measurement; search for the best solution through supervised learning with reference to a thoroughly-understood process from the problem definition, solution instantiation to measurement. 
Also, I elaborated BenchCouncil’s plan to define emerging and future computing challenges and collaborate with ComputerCouncil on open-source computer systems.




 

 \section{Acknowledgments}
I am very grateful to Mr. Shaopeng Dai for compiling the references, Mr. Shaopeng Dai and Mr. Qian He for drawing Figures 1, 2, 3, 4, and 6, and Dr. Lei Wang for discussing and contributing significantly to the presentations of Figures 1, 4, 6 and proofreading throughout this article.  Figure 5 is cited from~\cite{wang2022wpc} as an example with the permission of the authors.  Section 3 and a part of Section 5 are based on the unpublished technical report~\cite{zhan2019benchcouncil}, of which I am the lead author. The technical report~\cite{zhan2019benchcouncil} was based on my presentation at a BoF of SC 2019, and the web link is ~\url{https://www.benchcouncil.org/file/BenchCouncil-SC-BoF.pdf}. After the presentation, I drafted this article as the first author.  I am very grateful to the other authors for their discussions and contributions:  Dr. Lei Wang, Dr. Wanling Gao, and Dr. Rui Ren.

\printcredits

\bibliographystyle{cas-model2-names}

\bibliography{cas-refs}


\bio{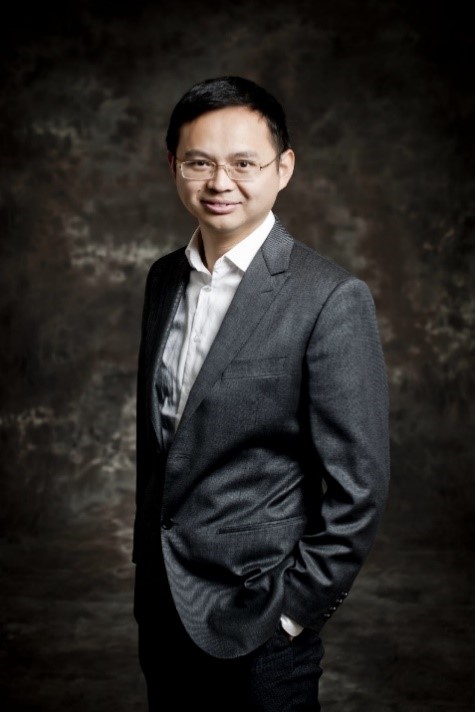}
Dr. Jianfeng Zhan is a Full Professor at Institute of Computing Technology (ICT), Chinese Academy of Sciences (CAS), and University of Chinese Academy of Sciences (UCAS), the director of Research Center for Advanced Computer Systems, ICT, CAS.  He received his B.E. in Civil Engineering and MSc in Solid Mechanics from Southwest Jiaotong University in 1996 and 1999, and his Ph.D. in Computer Science from Institute of Software, CAS, and UCAS in 2002. His research areas span from Chips, Systems to Benchmarks. A common thread is benchmarking, designing, implementing, and optimizing a diversity of systems. He has made substantial and effective efforts to transfer his academic research into advanced technology to impact general-purpose production systems. Several technical innovations and research results, including 35 patents, from his team, have been adopted in benchmarks, operating systems, and cluster and cloud system software with direct contributions to advancing the parallel and distributed systems in China or even in the world. He has supervised over ninety graduate students, post-doctors, and engineers in the past two decades. 
Dr. Jianfeng Zhan founds and chairs BenchCouncil and serves as the Co-EIC of TBench with Prof. Tony Hey. He has served as IEEE TPDS Associate Editor since 2018. He received the second-class Chinese National Technology Promotion Prize in 2006, the Distinguished Achievement Award of the Chinese Academy of Sciences in 2005, and the IISWC Best paper award in 2013, respectively. \endbio
\end{document}